%% file: skeleton.tex
\documentclass[a4paper,11pt]{article}
\usepackage{lineno}
\usepackage{comment}
\usepackage{xcolor}
\usepackage{xfrac}
\usepackage{amsmath}
\usepackage{amsfonts}
\usepackage{amsthm}
\usepackage{thmtools}
\usepackage{thm-restate}
\usepackage{amssymb}
\usepackage{slashed}
\usepackage{graphicx}
\usepackage{pos}
\usepackage{tikz}
\usepackage{tikz-feynman}
\tikzfeynmanset{warn luatex=false}
\usetikzlibrary{shapes.geometric, arrows, patterns, decorations.markings, arrows.meta}
\usetikzlibrary{calc}
\tikzstyle{top} = [rectangle, rounded corners, minimum width=3cm, minimum height=1cm, text centered, text width=3cm, draw=black, fill=red!30]
\allowdisplaybreaks

\input{pics}

\def\lgen{\mathcal{L}^{\text{gen}}}

\def\dij{\delta^{IJ}}
\def\zab{Z_{AB}^I}

\declaretheorem[name=Theorem]{theo}

\title{Operator Renormalization using Emergent Supersymmetries}

\author*[a]{Mrigankamauli Chakraborty}
\author[b]{Sven-Olaf Moch}

\affiliation[a,b]{II. Institute for Theoretical Physics, Hamburg University\\
Luruper Chaussee 149, D-22761 Hamburg, Germany}

\emailAdd{mrigankamauli.chakraborty@desy.de}
\emailAdd{sven-olaf.moch@desy.de}

\abstract{We develop a mechanism that enables supersymmetric Ward identities to be applied in non-supersymmetric theories. These identities are then used to streamline calculations in our target theories, potentially including phenomenological models. In these proceedings, we illustrate the method through operator renormalization in the Gross–Neveu–Yukawa model, where it leads to a significant optimization and a substantial reduction in computational effort. This serves as a toy example of the procedure that we ultimately aim to apply to Quantum Chromodynamics.}

\FullConference{17th International Symposium on Radiative Corrections: Applications of Quantum Field Theory to Phenomenology (RADCOR2025)\\
5-10 October 2025\\
Puri, India\\}

\begin{document}
\maketitle

\section{Introduction}\label{sec:intro}
\noindent 
The aim of our project is to use supersymmetry (SUSY) not as a phenomenological model, but as an optimization tool for phenomenological theories such as Quantum Chromodynamics (QCD) and the Standard Model. In this way, we set aside the lack of experimental evidence for SUSY and instead employ it purely as a computational aid to streamline calculations that are experimentally relevant.

The features of a supersymmetric theory that allow for computational optimization are the Ward identities generated by its super-Poincaré symmetry. These additional relations reduce the number of independent quantities that need to be computed. Such Ward identities include, for example, the equality of anomalous dimensions and $\beta$ functions, as well as possible non‑renormalization theorems. All of these appear as equations relating the bare 1PI Green's functions.

Similar identities do not exist in non‑SUSY theories, which include all phenomenologically relevant models. However, in these proceedings we introduce a formalism that connects non‑SUSY theories to SUSY theories in such a way that the Ward identities of the SUSY theory generate corresponding relations among the bare Green's functions of the non‑SUSY theory.

Such applications of SUSY to phenomenological models exist in the literature for limited cases.
For example, the gluonic sector of QCD is shared by $\mathcal{N}=4$ Super Yang-Mills (SYM) theory which in certain domains is exactly solvable.
A good review on this is provided in \cite{Henn:2020omi}.

Our formalism gives a more systematic way to connect QCD to all of $\mathcal{N}=1,2,4$ SYM theories, beyond just the gluonic sector.
These proceedings are based on the findings of \cite{emgsusy}, where we use the Gross-Neveu-Yukawa (GNY) model as a testing ground to develop this formalism.
In sec.~\ref{sec:lgen}, we motivate the definition of a generalised Lagrangian containing abstract flavour structures.
Then, using simple closure conditions on renormalization, we present a theorem that completely determines the value of the flavour factors from their abstract definition. 

The results for multiple theories, such as the GNY, Nambu-Jona-Lasinio-Yukawa (NJLY) and both the three- (3D) and four-dimensional (4D) Wess-Zumino models, can then be obtained from the results of the generalised Lagrangian by simply substituting for the respective number of scalars and fermions, cf. sec.~\ref{sec:emergentsusy}.
This not only gives us results for multiple theories from just one calculation, but the unification of both SUSY and non-SUSY models to one Lagrangian makes the SUSY Ward identities available also for the non-SUSY theories.

In sec.~\ref{sec:application} we discuss how this technique of making SUSY emerge out of a non-SUSY theory is then used to optimize calculations. The most dramatic effect of such an optimization can be seen in renormalization of twist-two operators.
The complicated numerator structures result in lengthy IBP reduction times, spanning days for the GNY model at three loops. In these proceedings, we demonstrate that our application of emergent SUSY reduces the computational time by 25\%.
In the case of QCD where such calculations take months, this kind of speed-up can save weeks to months of computation.

The anomalous dimensions of such operators in QCD are related to the parton
distribution functions via the Wilson OPE. Hence, this SUSY-based optimization tool has tremendous phenomenological relevance.
\section{The Generalised Lagrangian}\label{sec:lgen}
\noindent We begin by writing the most general form of a theory involving scalars and fermions, including a ternary Yukawa interaction and a quartic scalar coupling:
\begin{align}
\label{lgen}
 \mathcal{L}_{\text{gen}} &= \frac{1}{2}\partial_\mu \phi^I\partial^\mu \phi_I+ i\bar\psi^A\slashed{\partial}\psi_A-g(\bar Z^I)^{AB}\phi_I\psi_A\psi_B - g( Z^I)_{AB}\phi_I\bar\psi^A\bar\psi^B  - \frac{\lambda}{4!}M^{IJKL}\phi_I\phi_J\phi_K\phi_L
\, ,
\end{align}
where $\phi_I$ are real scalars and $\psi_A$ are Weyl fermions. The  scalar flavour index $I$ refers to the fundamental representation of a global $SO(n_s)$ flavour symmetry, while the fermion flavour index $A$ is for a fundamental global $SU(n_f)$ symmetry. Thus $(n_s,n_f)$ are respectively the number of scalars and fermions in the theory.

$(\bar Z^I)^{AB}, (Z^I)_{AB}$,  and $M^{IJKL}$ are \emph{flavour elements}, with 
no specific algebraic structure imposed at this stage. Their index structure is such that all flavour indices are fully contracted, ensuring that the Lagrangian transforms as a singlet under the flavour symmetry. Imposing the Hermiticity of the Lagrangian necessitates that the flavour elements $Z^I, \bar Z^I$ are symmetric and Hermitian (see \cite{emgsusy} for details.)

As stated in sec.~\ref{sec:intro}, we want to obtain the results of GNY, NJLY and Wess-Zumino models by just substituting the values for $(n_s,n_f)$ in the results of $\lgen$. Notice that none of those theories have any \emph{evanescent}\footnote{Evanescent terms in a Lagrangian are those that do not exist in the classical case, but are added by quantum corrections in order to cancel divergences from loop diagrams.} flavour terms. Hence, we must also impose that $\lgen$ does not have any evanescent flavour terms. Then, we can prove the following theorem:
\begin{theo}\label{thm:nonevanescence}
There are no evanescent flavour terms in the generalised Lagrangian $\lgen$ if and only if the $Z,\bar Z$ satisfy the following relation:
\begin{equation}\label{eq:flavclifford}
    Z^I\bar Z^J+Z^J\bar Z^I = 2\delta^{IJ}\;,
\end{equation}
i.e. the same relation satisfied by the Pauli matrices $\sigma^\mu,\bar\sigma^\mu$.
\end{theo}
We need to check non-evanescence only for primitive Feynman graphs (graphs without subdivergences). This means we only need to impose the following proportionality conditions:
\begin{equation}\label{lhsfer}
\begin{aligned}
 \olfer \propto \delta_A^{\ B}\quad\quad & \olsclro + \olsclrt \propto \dij
\,\quad\quad .\\
\olyuk \propto\; (Z^I)_{AB}\quad\quad & \tlyuk \qquad\propto\; (Z^I)_{AB}\\[3pt]
\olfouro + \olfourt +& \;(J\leftrightarrow K)\; +\; (J\leftrightarrow L)\;\propto\; M^{IJKL}\;.
\end{aligned}
\end{equation}
The above can be used to show that these conditions imply that eq.\eqref{eq:flavclifford} holds. We present a detailed proof of theorem \ref{thm:nonevanescence} in \cite{emgsusy}. We then have, for the quartic flavour factor:
\begin{equation}\label{mijkl}
    M^{IJKL} = (\delta^{IJ}\delta^{KL}+\delta^{IK}\delta^{JL}+\delta^{IL}\delta^{JK})
\end{equation}
where the proportionality constant is chosen so that the Feynman rule for the quartic vertex is $-i\lambda M^{IJKL}$.

These conditions on the flavour elements $\zab$ and $M^{IJKL}$ completely determine the bare 1PI correlators for the propagators and vertices as functions of $(n_s,n_f)$.
This is possible because the eq.~\eqref{eq:flavclifford} can be used to calculate the traces of $Z,\bar Z$ in a similar way as Dirac traces. Detailed derivations of the relevant trace identities can be found in \cite{emgsusy}.

\section{Emergent SUSY in $\lgen$}
\label{sec:emergentsusy}
To demonstrate how SUSY emerges from the generalised Lagrangian, we calculate the fermion and scalar propagators and the Yukawa vertex up to four loops with \texttt{FORCER}~\cite{Ruijl:2017cxj}. We use \texttt{Qgraf}~\cite{Nogueira:1991ex} to generate the diagrams. 
The \texttt{Qgraf} output is converted into a \texttt{FORCER} compatible input with the computer algebra system \texttt{FORM}~\cite{Davies:2026cci}, which we use to implement the Feynman rules, identify topologies, and carry out the flavour and spinor traces.
The loop integrations are then carried out with \texttt{FORCER}.
The renormalization of the bare quantities in the $\overline{\rm MS}$-scheme proceeds in the standard way.

\begin{figure}
    \centering
    \includegraphics[width=0.5\linewidth]{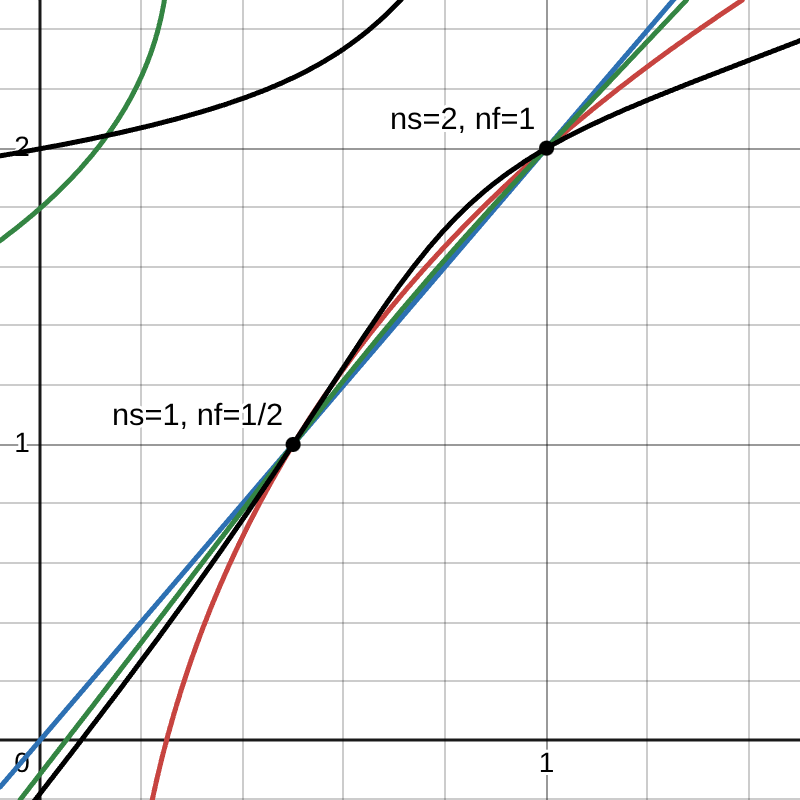}
    \caption{Plot of the SUSY Ward Identity $\gamma_f=\gamma_s$ at every loop order up to 4 loops, y-axis is $n_s$, x-axis is $n_f$. Curves for every loop order intersect at the same 2 points, signifying that $\lgen$ as a whole becomes supersymmetric at those two points.}
    \label{fig:emgsusytheo}
\end{figure}

The anomalous dimensions and $\beta$ functions thus calculated for $\lgen$ are functions of $n_s,n_f$. We say SUSY emerges from $\lgen$ if at certain points $(n_s,n_f)$ SUSY Ward identities hold. To that purpose, consider the Ward identity
\begin{equation}
\label{eq:gf=gs}
  \gamma_f = \gamma_s  \, ,
\end{equation}
where $\gamma_f$ and $\gamma_s$ denote the anomalous dimensions of the fermionic and scalar fields, respectively.
At every loop order, we plot the curves in a $(n_f, n_s)$ plane for the above equation.
SUSY then emerges at the point where all these curves intersect.
Fig.\ref{fig:emgsusytheo} shows the curves up to four loops obtained from our calculations, exhibiting emergent SUSY at two points: $(2,1)$ and $(1, \dfrac{1}{2})$.

Since we employ Weyl fermions, the value $n_f = 1$ corresponds to two fermionic degrees of freedom.
Consequently, the SUSY point $(2,1)$ represents a theory with two scalar and two fermionic degrees of freedom, precisely matching the field content of the 4D Wess-Zumino model.
The quantities also match the values calcuated for the Wess-Zumino model in the literature \cite{Abbott:1980jk,Sen:1981hk}, signifying that this point refers to the 4D Wess-Zumino model. 

The point $\left(1, \tfrac{1}{2}\right)$ corresponds to a theory with a single fermionic degree of freedom, which is not possible in four dimensions. 
The paricle content coincides with that of the 3D WZ model~\cite{Grover:2013rc}, a known emergent SUSY of the GNY model.

There is also another observation: the emergent SUSY curve for the highest transcendental number at any loop order is always the same as the first loop order SUSY curve
\begin{equation}
    n_s-2n_f=0\;,
\end{equation}
which refers to the equality of bosonic and fermionic degrees of freedom. This has implications for SYM theories which we will explore in a future project.

\section{Using emergent SUSY to optimize calculations in GNY}\label{sec:application}

\noindent The results for the GNY model \cite{Mihaila:2017ble,Zerf:2017zqi} can also be obtained from the results of $\lgen$ by setting $n_s=1$.
This allows us to apply the SUSY Ward identities to GNY even though it is not supersymmetric. 
To demonstrate how this is done, consider the two‑loop propagator 1PI correlators for the fermion (solid lines) and the scalar fields (dashed lines) obtained from $\lgen$:
\begin{equation}\label{eq:emgsusytwoloop}
    \begin{aligned}
        \Gamma_\psi^{(2)} &= a_0^2n_s(n_s-2)\,\intpsio +a_0^2n_fn_s\,\intpsitw \\
        \Gamma_\phi^{(2)} &= a_0^2n_f(n_s-2)\,\intphio+a_0^2n_fn_s\,\intphitw + \lambda_0^2(n_s+2)\,\intphithr\;.
    \end{aligned}
\end{equation}
The diagrams displayed above are evaluated with the appropriate Born projections, such that the corresponding tree‑level contributions are normalised to unity.
At this point, the integrals are yet not calculated.
Then, substituting for the SUSY points gives us the following equations among the integrals on account of the Ward identities:
\begin{equation}
\label{eq:optimizing-susy-1}
    \begin{aligned}
         -1\,\intpsio + \frac{1}{2}\intpsitw&= -\frac{1}{2}\intphio + \frac{1}{2}\intphitw+3\intphithr\\
        2\intpsitw &=2\intphitw+4\intphithr \; .
    \end{aligned}
\end{equation}
The optimization happens here: we can choose to eliminate two of the integrals using the above Ward identities. This not only eliminates the integrals for the SUSY points, but also for the generalised Lagrangian as a whole.
Hence, we need to carry out fewer integrations when we calculate results of $\lgen$ itself, and then obtain the results of the GNY model by substituting $n_s=1$ with the cost of almost no extra computational overhead.

At higher loops, the most non-trivial topologies often satisfy the SUSY Ward identities independently of other topologies. These extra equations allow further eliminations beyond the two obtained from the SUSY points alone. For example, at three loops we have the following independent relation for the non-planar topology:
\begin{equation}
    \nonpphi=\quad2\; \nonppsi
\end{equation}
The advantage of this optimization method becomes substantial in the context of operator renormalization, where it can reduce computational time at the order of days and weeks.
Consider the following twist-two flavour singlet operators in the GNY model:
\begin{align}   
\mathcal{O}^{\mu_1\mu_2\ldots\mu_N}_\psi(N) &= \frac{(-i)^{N-1}}{N!}\bar\psi^A\gamma^{[\mu_1}\partial^{\mu_2\ldots\mu_N]}\psi_A\:-\:\text{traces}\\
    \mathcal{O}^{\mu_1\mu_2\ldots\mu_N}_\phi(N) &= \frac{(-i)^{N}}{N!}\phi\,\partial^{[\mu_1\ldots\mu_N]}\,\phi\:-\:\text{traces}
\end{align}
where $N$ denotes the spin, which can only take even values for the scalar operator.
The above operators mix under renormalization and their anomalous dimensions are represented by the matrix
\begin{equation}\label{oper_ren}
    \begin{bmatrix}
        \gamma_{\psi\psi} & \gamma_{\phi\psi} \\
        \gamma_{\psi\phi} & \gamma_{\phi\phi}
    \end{bmatrix}
\, .
\end{equation}
In our work~\cite{emgsusy}, we show that for the Wess–Zumino model the following Ward identity holds for all even values of N and at all loop orders:
\begin{equation}\label{OME_gm}
        2\gamma_{\psi\psi}+4\gamma_{\phi\psi} = 2\gamma_{\phi\phi}+\gamma_{\psi\phi} 
        \, .
    \end{equation}
We have verified via explicit calculation up to three loops that the above Ward identity emerges at the same two emergent SUSY points of $\lgen$.
For the 3D Wess–Zumino point, we need to perform the Clifford algebra in a manner that is fully consistent with three dimensions. Details given in \cite{emgsusy}.

Obtaining the general-$N$ expression for operator anomalous dimensions is a topic of great interest, especially in QCD, since it allows one to determine the splitting functions that govern the evolution of parton distribution functions via the inverse Mellin transform. 
The usual approach is to compute the anomalous dimensions individually up to sufficiently high values of the Mellin moment $N$, and then infer the general expression using the LLL algorithm, see e.g.~\cite{Kniehl:2025ttz}. 
However, at higher values of $N$, the complicated numerator structure leads to long computational times with \texttt{FORCER}. For example, in the GNY model, calculating all even moments at three loops up to $N=30$ takes approximately 14 days. The method of emergent SUSY creates a substantial speed up here. By eliminating a certain number of non-planar integrals, the computational time is reduced by 25\% at every value of $N$, which is a significant improvement.

\section{Summary and future outlook}
\noindent In these proceedings we discussed the technique of the generalised Lagrangian that can unify SUSY theories with useful (and possibly phenomenological) non-SUSY models.
This allows us to ``import" SUSY Ward identities into non-SUSY models, thereby allowing us to use optimization tricks hitherto not available for non-SUSY theories.

We showcase our method for the case of the GNY model here.
Using the emergent SUSY of the generalised Lagrangian, we were able to incorporate SUSY Ward identities from two SUSY models (3D and 4D Wess-Zumino) into the 4D GNY model. We then applied it to the computation of operator anomalous dimensions, where a substantial speedup led to significant savings in computation time.

Our ultimate objective for developing this method is to apply emergent SUSY to QCD, where it could lead to saving months of computation.
We already have the generalised Lagrangian to unify QCD with SYM theories (see \cite{Chakraborty:2026hdv}). There are a few more subtle steps that we need due to the presence of the gauge symmetry, and this would be shown in a future work. 

\providecommand{\href}[2]{#2}\begingroup\raggedright\endgroup

\end{document}

%% file: pics.tex
\newcommand{\olsclro}{\begin{minipage}{2.5cm}
\begin{tikzpicture}
\draw (-1, 0.2) node{${\scriptstyle I}$};
\draw (1, 0.2) node{${\scriptstyle J}$};
        \begin{feynman}
             \vertex (b) at (0.5,0);
             \vertex (d) at (-0.5,0);
             \vertex (e) at (-1, 0);
             \vertex (f) at (1, 0);
             \diagram* {
             (b) -- [fermion, half left] (d);
             (b) -- [fermion, half right] (d);
             (e) -- [scalar] (d);
             (f) -- [scalar] (b); 
             };
          \end{feynman}
\end{tikzpicture}
\end{minipage}}

\newcommand{\olsclrt}{\begin{minipage}{2.5cm}
\begin{tikzpicture}
\draw (-1, 0.2) node{${\scriptstyle I}$};
\draw (1, 0.2) node{${\scriptstyle J}$};
        \begin{feynman}
             \vertex (b) at (0.5,0);
             \vertex (d) at (-0.5,0);
             \vertex (e) at (-1, 0);
             \vertex (f) at (1, 0);
             \diagram* {
             (d) -- [fermion, half left] (b);
             (d) -- [fermion, half right] (b);
             (e) -- [scalar] (d);
             (f) -- [scalar] (b); 
             };
          \end{feynman}
\end{tikzpicture}
\end{minipage}}

\newcommand{\olfer}{\begin{minipage}{2.5cm}
\begin{tikzpicture}
\draw (-1, -0.2) node{${\scriptstyle A}$};
\draw (1, 0.2) node{${\scriptstyle B}$};
        \begin{feynman}
             \vertex (b) at (0.5,0);
             \vertex (d) at (-0.5,0);
             \vertex (e) at (-1, 0);
             \vertex (f) at (1, 0);
             \diagram* {
             (b) -- [scalar, half left] (d);
             (d) -- [fermion, half left] (b);
             (d) -- [fermion] (e);
             (f) -- [fermion] (b); 
             };
          \end{feynman}
\end{tikzpicture}
\end{minipage}}

\newcommand{\olyuk}{\begin{minipage}{2cm}
\begin{tikzpicture}
\draw (0, 1.45) node{${\scriptstyle I}$};
\draw (-0.9, -0.2) node{${\scriptstyle A}$};
\draw (0.9, -0.2) node{${\scriptstyle B}$};
        \begin{feynman}
             \vertex (b) at (0.5,0);
             \vertex (c) at (-0.5,0);
             \vertex (a) at (0,0.85);
             \vertex (d) at (0, 1.25);
             \vertex (e) at (-0.75, -0.42);
             \vertex (f) at (0.75, -0.42);
             \diagram* {
             (b) -- [scalar] (c);
             (a) -- [fermion] (b);
             (a) -- [fermion] (c);
             (e) -- [fermion] (c);
             (f) -- [fermion] (b);
             (a) -- [scalar] (d);
             };
          \end{feynman}
\end{tikzpicture}
\end{minipage}}

\newcommand{\tlyuk}{\begin{minipage}{2cm}
\begin{tikzpicture}
\draw (0, 1.45) node{${\scriptstyle I}$};
\draw (-1.65, -0.62) node{${\scriptstyle A}$};
\draw (1.65, -0.62) node{${\scriptstyle B}$};
        \begin{feynman}
             \vertex (b) at (0.5,0.21);
             \vertex (c) at (-0.5,0.21);
             \vertex (g) at (1,-0.42);
             \vertex (h) at (-1,-0.42);
             \vertex (a) at (0,0.84);
             \vertex (d) at (0, 1.25);
             \vertex (e) at (-1.5, -0.84);
             \vertex (f) at (1.5, -0.84);
             \diagram* {
             (b) -- [scalar] (h);
             (c) -- [scalar] (g);
             (b) -- [fermion] (g);
             (c) -- [fermion] (h);
             (b) -- [fermion] (a);
             (c) -- [fermion] (a);
             (e) -- [fermion] (h);
             (f) -- [fermion] (g);
             (a) -- [scalar] (d);
             };
          \end{feynman}
\end{tikzpicture}
\end{minipage}}

\newcommand{\olfouro}{\begin{minipage}{2cm}
\begin{tikzpicture}
\draw (-0.8, 0.5) node{${\scriptstyle I}$};
\draw (0.8, 0.5) node{${\scriptstyle J}$};
\draw (0.8, -1.1) node{${\scriptstyle K}$};
\draw (-0.8, -1.1) node{${\scriptstyle L}$};
        \begin{feynman}
             \vertex (a) at (0.5,0.17);
             \vertex (d) at (-0.5,0.17);
             \vertex (b) at (0.5,-0.83);
             \vertex (c) at (-0.5,-0.83);
             \vertex (g) at (0.75, 0.42);
             \vertex (h) at (-0.75, 0.42);
             \vertex (e) at (-0.75, -1.08);
             \vertex (f) at (0.75, -1.08);
             \diagram* {
             (g) -- [scalar] (a);
             (h) -- [scalar] (d);
             (e) -- [scalar] (c);
             (f) -- [scalar] (b);
             (a) -- [fermion] (d);
             (a) -- [fermion] (b);
             (c) -- [fermion] (b);
             (c) -- [fermion] (d);
             };
          \end{feynman}
\end{tikzpicture}
\end{minipage}}

\newcommand{\olfourt}{\begin{minipage}{2cm}
\begin{tikzpicture}
\draw (-0.8, 0.5) node{${\scriptstyle I}$};
\draw (0.8, 0.5) node{${\scriptstyle J}$};
\draw (0.8, -1.1) node{${\scriptstyle K}$};
\draw (-0.8, -1.1) node{${\scriptstyle L}$};
        \begin{feynman}
             \vertex (a) at (0.5,0.17);
             \vertex (d) at (-0.5,0.17);
             \vertex (b) at (0.5,-0.83);
             \vertex (c) at (-0.5,-0.83);
             \vertex (g) at (0.75, 0.42);
             \vertex (h) at (-0.75, 0.42);
             \vertex (e) at (-0.75, -1.08);
             \vertex (f) at (0.75, -1.08);
             \diagram* {
             (g) -- [scalar] (a);
             (h) -- [scalar] (d);
             (e) -- [scalar] (c);
             (f) -- [scalar] (b);
             (d) -- [fermion] (a);
             (b) -- [fermion] (a);
             (b) -- [fermion] (c);
             (d) -- [fermion] (c);
             };
          \end{feynman}
\end{tikzpicture}
\end{minipage}}

\newcommand{\intpsio}{
\begin{minipage}{1.8cm}
\begin{tikzpicture}[scale=1.5]
    \begin{feynman}
             \vertex (vg1) at (-0.4,0);
             \vertex (vg2) at (0.4,0);
             \vertex (vq2) at (0,0.4);
             \vertex (vs) at (0,-0.4);
             \vertex (d) at (0.6,0);
             \vertex (e) at (-0.6,0);
             \diagram* {
             (e) --   (vg1);
             (d) --   (vg2);
             (vq2) --   (vg1);
             (vs) --   (vg2);
             (vs) -- (vq2);
             (vg1) -- [dashed, dash pattern=on 1.5pt off 1.5pt] (vs);
            (vq2) -- [dashed, dash pattern=on 1.5pt off 1.5pt] (vg2);
             };
    \end{feynman}
\end{tikzpicture}
    \end{minipage}
}

\newcommand{\intphio}{
\begin{minipage}{1.8cm}
\begin{tikzpicture}[scale=1.5]
    \begin{feynman}
             \vertex (vg1) at (-0.4,0);
             \vertex (vg2) at (0.4,0);
             \vertex (vq2) at (0,0.4);
             \vertex (vs) at (0,-0.4);
             \vertex (d) at (0.6,0);
             \vertex (e) at (-0.6,0);
             \diagram* {
             (vs) --   (vg1);
             (vq2) --   (vg2);
             (vq2) --   (vg1);
             (vs) --   (vg2);
             (vs) -- [dashed, dash pattern=on 1.5pt off 1.5pt] (vq2);
             (vg1) -- [dashed, dash pattern=on 1.5pt off 1.5pt] (e);
             (d) -- [dashed, dash pattern=on 1.5pt off 1.5pt] (vg2);
             };
    \end{feynman}
\end{tikzpicture}
    \end{minipage}
}

\newcommand{\intpsitw}{
\begin{minipage}{1.8cm}
\begin{tikzpicture}[scale=1.5]
\draw (0,0.5) node{};
\draw (0.4,0) arc(0:180:0.4);
\draw (0.4,0) [dashed, dash pattern=on 1.5pt off 1.5pt] arc(360:300:0.4);
\draw (-0.4,0) [dashed, dash pattern=on 1.5pt off 1.5pt] arc(180:240:0.4);
    \begin{feynman}
             \vertex (vg1) at (-0.4,0);
             \vertex (vg2) at (0.4,0);
             \vertex (vq2) at (-0.2,-0.34);
             \vertex (vq1) at (0.2,-0.34);
             \vertex (d) at (0.6,0);
             \vertex (e) at (-0.6,0);
             \diagram* {
             (e) --   (vg1);
             (d) --   (vg2);
             (vq2) -- [half left] (vq1);
             (vq2) -- [half right] (vq1);
             };
    \end{feynman}
\end{tikzpicture}
    \end{minipage}
}

\newcommand{\intphitw}{
\begin{minipage}{1.8cm}
\begin{tikzpicture}[scale=1.5]
\draw (0,0.5) node{};
\draw (0.4,0) arc(0:180:0.4);
\draw (0.4,0) arc(360:300:0.4);
\draw (-0.4,0)  arc(180:240:0.4);
\draw (0.4,0)  arc(360:300:0.4);
    \begin{feynman}
             \vertex (vg1) at (-0.4,0);
             \vertex (vg2) at (0.4,0);
             \vertex (vq2) at (-0.2,-0.34);
             \vertex (vq1) at (0.2,-0.34);
             \vertex (d) at (0.6,0);
             \vertex (e) at (-0.6,0);
             \diagram* {
             (e) --  [dashed, dash pattern=on 1.5pt off 1.5pt]  (vg1);
             (d) --   [dashed, dash pattern=on 1.5pt off 1.5pt] (vg2);
             (vq2) -- [half left] (vq1);
             (vq2) -- [dashed, dash pattern=on 1.5pt off 1.5pt, half right] (vq1);
             };
    \end{feynman}
\end{tikzpicture}
    \end{minipage}
}

\newcommand{\intphithr}{
\begin{minipage}{1.8cm}
\begin{tikzpicture}[scale=1.5]
\draw (0.4,0) [dashed, dash pattern=on 1.5pt off 1.5pt] arc(0:360:0.4);
    \begin{feynman}
             \vertex (vg1) at (-0.4,0);
             \vertex (vg2) at (0.4,0);
             \vertex (vq2) at (-0.2,-0.34);
             \vertex (vq1) at (0.2,-0.34);
             \vertex (d) at (0.6,0);
             \vertex (e) at (-0.6,0);
             \diagram* {
             (vg1) --   [dashed, dash pattern=on 1.5pt off 1.5pt] (vg2);
             (e) --  [dashed, dash pattern=on 1.5pt off 1.5pt]  (vg1);
             (d) --   [dashed, dash pattern=on 1.5pt off 1.5pt] (vg2);
             };
    \end{feynman}
\end{tikzpicture}
    \end{minipage}
}

\newcommand{\nonpphi}{
\begin{minipage}{3.5cm}
\begin{tikzpicture}[scale=2]
    \begin{feynman}
             \vertex (vg1) at (-0.6,0);
             \vertex (vg2) at (0.6,0);
             \vertex (vq1) at (-0.3,0.3);
             \vertex (vq2) at (-0.3,-0.3);
             \vertex (vq3) at (0.3,-0.3);
             \vertex (vq4) at (0.3,0.3);
             \vertex (v1) at (0.05,-0.05);
             \vertex (v2) at (-0.05,0.05);
             \vertex (d) at (0.8,0);
             \vertex (e) at (-0.8,0);
             \diagram* {
             (vq1) --   (vg1);
             (vq3) --   (vg2);
             (vq2) --   (vg1);
             (vq4) --   (vg2);
             (vq3) --   (vq2);
             (vq4) --   (vq1);
             (vq1) -- [dashed, dash pattern=on 1.5pt off 1.5pt] (v2);
             (v1) -- [dashed, dash pattern=on 1.5pt off 1.5pt] (vq3);
             (vq4) -- [dashed, dash pattern=on 1.5pt off 1.5pt] (vq2);
             (vg1) -- [dashed, dash pattern=on 1.5pt off 1.5pt] (e);
             (d) -- [dashed, dash pattern=on 1.5pt off 1.5pt] (vg2);
             };
    \end{feynman}
\end{tikzpicture}
    \end{minipage}
}

\newcommand{\nonppsi}{
\begin{minipage}{3.5cm}
\begin{tikzpicture}[scale=2]
    \begin{feynman}
             \vertex (vg1) at (-0.6,0);
             \vertex (vg2) at (0.6,0);
             \vertex (vq1) at (-0.3,0.3);
             \vertex (vq2) at (-0.3,-0.3);
             \vertex (vq3) at (0.3,-0.3);
             \vertex (vq4) at (0.3,0.3);
             \vertex (v1) at (0.05,-0.05);
             \vertex (v2) at (-0.05,0.05);
             \vertex (d) at (0.8,0);
             \vertex (e) at (-0.8,0);
             \diagram* {
             (vq1) --   (vg1);
             (vq3) --   (vg2);
             (vq2) -- [dashed, dash pattern=on 1.5pt off 1.5pt]  (vg1);
             (vq4) --  [dashed, dash pattern=on 1.5pt off 1.5pt] (vg2);
             (vq3) --   (vq2);
             (vq4) --   (vq1);
             (vq1) -- [dashed, dash pattern=on 1.5pt off 1.5pt] (v2);
             (v1) -- [dashed, dash pattern=on 1.5pt off 1.5pt] (vq3);
             (vq4) -- (vq2);
             (vg1) --  (e);
             (d) -- (vg2);
             };
    \end{feynman}
\end{tikzpicture}
    \end{minipage}
}